\documentclass{aa}

\usepackage{graphicx}
\usepackage{txfonts}
\usepackage{amstext}
\usepackage{upgreek}

\def\PGPU{$\varphi-$GPU }

\def\ba{\mbox{\boldmath $a$}}
\def\acck#1{\ba_i^{(#1)}}
\def\absak#1{|\acck{#1}|}

\renewcommand{\arraystretch}{1.8}

\def\gapprox{\;\rlap{\lower 3.0pt                       
        \hbox{$\sim$}}\raise 2.5pt\hbox{$>$}\;}
\def\lapprox{\;\rlap{\lower 3.1pt                       
        \hbox{$\sim$}}\raise 2.7pt\hbox{$<$}\;}

\newcommand{\be}{ \begin{equation} }
\newcommand{\ee}{\end{equation}}

\newcommand{\ben}{\begin{enumerate}}
\newcommand{\een}{\end{enumerate}}

\makeatletter
\renewcommand*\aa@pageof{, page \thepage{} of \pageref*{LastPage}}
\makeatother
\newcommand{\orcid}[1]{\href{https://orcid.org/#1}{\protect\includegraphics[width=8pt]{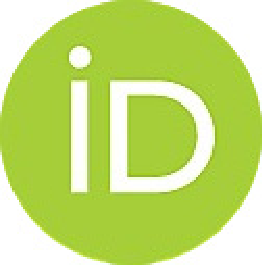}}}
\usepackage{hyperref}
\usepackage{ulem}
\hypersetup{colorlinks = true, citecolor = {blue}, urlcolor= {blue}, linkcolor={red}}

\begin{document}


\title{Dynamical evolution timescales for the triple supermassive black hole     system in NGC~6240}

\titlerunning{Triple SMBH system in NGC~6240}
\author{P.~Berczik\orcid{0000-0003-4176-152X}\inst{1,2,3,4}
\and
M.~Sobolenko\orcid{0000-0003-0553-7301}\inst{1,2}
\and
M.~Ishchenko\orcid{0000-0002-6961-8170}\inst{1,4}
}

\institute{Main Astronomical Observatory, National 
Academy of Sciences of Ukraine,
27 Akademika Zabolotnoho St, 03143 Kyiv, Ukraine, 
\email{\href{mailto:berczik@mao.kiev.ua}{berczik@mao.kiev.ua}}
\and
Nicolaus Copernicus Astronomical Centre Polish Academy of Sciences, ul. Bartycka 18, 00-716 Warsaw, Poland
\and
Konkoly Observatory, Research Centre for Astronomy and Earth Sciences, HUN-REN CSFK, MTA Centre of Excellence, Konkoly Thege Mikl\'os \'ut 15-17, 1121 Budapest, Hungary
\and
           Fesenkov Astrophysical Institute, Observatory 23, 050020 Almaty, Kazakhstan
}
   
\date{Received xxx / Accepted xxx}

\abstract    
{}
{Based on the available observational data from the literature, we analysed the dynamics of the NGC~6240 galaxy central supermassive black hole (SMBH) system.}
{For the dynamical modelling of this triple SBMH system, we used the massively parallel and GPU accelerated \PGPU direct summation $N$-body code. Following a long-timescale modelling of the triple system, we carried out a very detailed time output analysis of the  von~Zeipel-Lidov-Kozai (ZLK) oscillations for the black holes.}
{According to our {Newtonian} simulation results, for all models and randomisations, the bound system from {\tt S1+S2} components formed at $\approx3.6$~Myr. The formation of the bound hierarchical triple system {\tt S+N} occurred at $\approx18$~Myr. Over the course of {these Newtonian} simulations of the evolution of the triple SMBH\ system and the surrounding environment in NGC~6240,  ZLK oscillations were detected (in most cases) for the binary components. The inclination angle between the orbital angular momentum of binary components aptly coincides with the theoretical calculations of the ZLK mechanism.}
{In our set of randomised 15 Newtonian $N$-body dynamical galaxy models in 13 systems, we were able to detect a ZLK mechanism. In contrast, our extra few-body post-Newtonian runs (for one randomisation case) show it is only for the large inner binary initial eccentricity (in our case $\gtrsim$0.9) that we are able to observe the  possibility of the inner binary merging, due to the post-Newtonian energy radiation effects. For the lower eccentricity cases, the test runs show no sign of possible merging or any~ZLK oscillations in the system.}

     
\titlerunning{Triple SMBH system in NGC~6240}
\authorrunning{P.~Berczik et al.}
\maketitle

\section{Introduction}\label{sec:Intr}

Our understanding today of the co-evolution of galaxies and their supermassive black holes \citep[SMBHs;][]{Kormendy2013} has been well established. In nearly  all nearby spatially resolved galactic centres, an SMBH has been detected. Such a tight connection between SMBHs and their host galaxies sheds light on the mutual cosmological and dynamical existence of these components. The observed (and numerically modelled) galactic mergers inevitably lead us to the presence of SMBH mergers \citep{Begelman1980, KH2000, HK2000}. {In  previously proposed models of SMBH mergers, there has been a well known `final parsec problem' reported, as described in \cite{Berczik2005} and \cite{Milosavljevic2003, Milosavljevic2003b}. A possible solution to this problem was originally found in the non-axisymmetric (or even triaxial) dynamical loss cone repopulation mechanisms \citep{Berczik2006, Nixon2011, Preto2011, Khan2013, Gualandris2022, Berczik2022, Zhu2024}}. The typical timescales of such mergers cover billions of years, whereby new galactic mergers may occur concurrently to  an unfinished prior merger. Such a combination of events may result in the presence of three (or more) SMBHs  \citep{Valt1994, HL2007, Ryu2018}.  

The study of the triple SMBH phase of merging galaxies is a complex and  time-consuming dynamic problem even today. We know of several very significant studies devoted to this topic, namely, those of \cite{ME1994}, \cite{HL2007}, \cite{Pau2010} and \cite{Koehn2023}. As a general conclusion from these papers, we can assume that the triple SMBH stage is quite common over galactic merger history. Generally, as a result of such a triple dynamical evolution, we have two outcomes: ejection or coalescence {\citep{Bonetti2016, Bonetti2019, Mannerkoski2021}}. {Although the triple interaction could potentially lead to the ejection of one or even all of the
SMBHs \citep{Valt1994}, most systems are long-lived \citep[$\approx$1~Gyr;][]{HL2007}; thus, a final coalescence is more common than ejection, confirming the analytical results of \cite{ME1996}}. Based on the reports cited above, we can conclude that the coalescence scenario is more favourable compared with the ejection scenario.  

The NGC~6240 galaxy ($z=0.0243$, $D_{\rm L}=111.2$~Mpc\footnote{\url{https://ned.ipac.caltech.edu/}}) is one of the firstly directly detected dual active galactic nuclei (AGNs), which shows the signature of the galactic merging \citep{Komossa2003,Puccetti2016}. 
The galaxy centre shows intensive activity across different spectral ranges: the presence of H$_{2}$O maser \citep{Hagiwara2002,Hagiwara2003}, ultra-luminous infrared emission \citep{Gerssen2004,Sanders2003,Iono2007}, and X-ray emission \citep{Pasquali2003,Fabbiano2020}. The dual-AGN nature of the NGC~6240 nuclei has been seen in the extensive Multi-Element Radio Linked Interferometer Network observations at 1.4~Ghz and 5~Ghz by \cite{Beswick2001}. The detected two main radio sources: north ({\tt N}) and south ({\tt S}), which are also well matched with the compact X-ray sources. The {\tt N} nucleus can be clearly classified as an AGN according to the characteristics in the radio band observations. The {\tt S} nucleus spectrum contains emission from the AGN and circumnuclear starburst and supernova remnants \citep{Gallimore2004,Hagiwara2011}. Recently, the results from \cite{Kollatschny2020} and \cite{Fabbiano2020} on the double structure of the {\tt S} nucleus have further contributed to the discussion.

The SMBH mass in the {\tt S} nuclei falls within the range of $(0.87-2.0)\times10^9\;\rm M_\odot$, based on high-resolution stellar kinematic results from \citep{Medling2011}. Using K-band data from the Very Large Telescope and the classical SMBH-$\sigma$ relation proposed by \citep{Tremaine2002}, the SMBH masses for the {\tt N} and {\tt S} nuclei were estimated as $(1.4\pm0.4)\times10^8\rm\;M_\odot$ and $(2.0\pm0.4)\times10^8\rm\;M_\odot$, respectively \citep{Engel2010}. \cite{Engel2010} studied the motion of molecular gas through CO emission and associated it with circular movements. However, \cite{Treister2020} later linked it to turbulence motion. More recently, the velocity dispersion obtained with the MUSE instrument was found to correspond to an {\tt N} nucleus black hole mass of $(3.6\pm0.8)\times10^8\rm\;M_\odot$ and a combined {\tt S = S1+S2} nucleus black hole mass of $(8.0\pm0.8)\times10^8\rm\;M_\odot$ \citep{Kollatschny2020}.

This object has already been investigated by our team in a few papers \citep{Sobolenko2021, Sobolenko2022}. In both papers, we restricted our numerical modelling to the binary SMBH case. Based on our extensive post-Newtonian study, we subsequently estimated maximum merging timescale as $\approx70$~Myr for {\tt N} and {\tt S} SMBH components \citep[see Fig.~5 in][]{Sobolenko2022}. 

{The paper is organised as follows. In Sect.~\ref{sec:num} we introduce the initial conditions for our physical and numerical model of the NGC~6240 triple SMBH system and shortly describe the dynamical integration procedure. In Sect.~\ref{sec:triple}, we present the dynamical evolution of our SMBH triple system with von~Zeipel-Lidov-Kozai (ZLK) oscillations. In Sect.~\ref{sec:conc}, we present the results of our numerical model integration and summarise our findings.}

\section{Triple SMBH initial conditions}\label{sec:num}
\subsection{Physical model}\label{sec:phys-mod}
To build the initial physical model, the estimated parameters of the central black holes, along with the stellar and gas components, were collected based on existing observational data \citep{Kollatschny2020, Medling2011, Engel2010, Tecza2000, Tacconi1999}. {The selected physical parameters which were adopted in our numerical models are presented in Table~\ref{tab:init-par}. The values for the galaxy components stellar mass $M_{\ast}$ ({\tt S1, S2, N}) were selected from the \cite[see In Sect. 4.2 on page 10,][]{Kollatschny2020}. These masses were derived from the components dynamical Jeans modelling using the corresponding stellar velocity dispersion maps \cite[see Fig. 10,][]{Kollatschny2020}.}

\begin{table}
\caption{NGC~6240 parameters of the physical model.}
\label{tab:init-par}
\centering
\renewcommand{\arraystretch}{1.2}
\resizebox{0.49\textwidth}{!}{
\begin{tabular}{cccccccc}
\hline
\hline
Nucleus & $M_{\ast}$ & $M_{\bullet}$ & $M_{\rm tot}$  &  $q$ & 
$\Delta R$ & $r$ & $a$ \\
 & $10^{10}\rm\;M_{\odot}$ & $10^{8}\rm\;M_{\odot}$ & $10^{10}\rm\;M_{\odot}$ &  & pc & pc  & pc \\
(1) & (2) & (3) & (4) & (5) & (6) & (7) & (8) \\
\hline
\hline
{\tt S1} & 1.23 & 7.1 & 1.301 & 0.058 & 198 & 250 & 25.0 \\
{\tt S2} & 0.07 & 0.9 & 0.079 & 0.129 &     & 100 & 10.0 \\
{\tt N}  & 0.25 & 3.6 & 0.286 & 0.144 & {856} & 250 & 25.0 \\
\hline
{\tt B}  & 0.21 &     & 0.210 &       &     & 856 & 85.6 \\
\hline
\end{tabular}
}
\begin{minipage}{\linewidth}
\smallskip
\textit{Note:}
(1) nuclei ID, (2) stellar mass, (3) BH mass, (4) summarised mass of stars and BH $M_{\rm tot}=M_{\ast}+M_{\bullet}$, (5) mass ratio BH to stars of $q=M_{\bullet}/M_{\ast}$, (6) initial separation for SMBHs ({\tt S1},{\tt S2}) and ({\tt S},{\tt N}) pairs, (7) {cutting} radius of the Plummer sphere, and (8) Plummer radius. 
\end{minipage}
\end{table}

\begin{figure}
\centering
\includegraphics[width=0.99\linewidth]{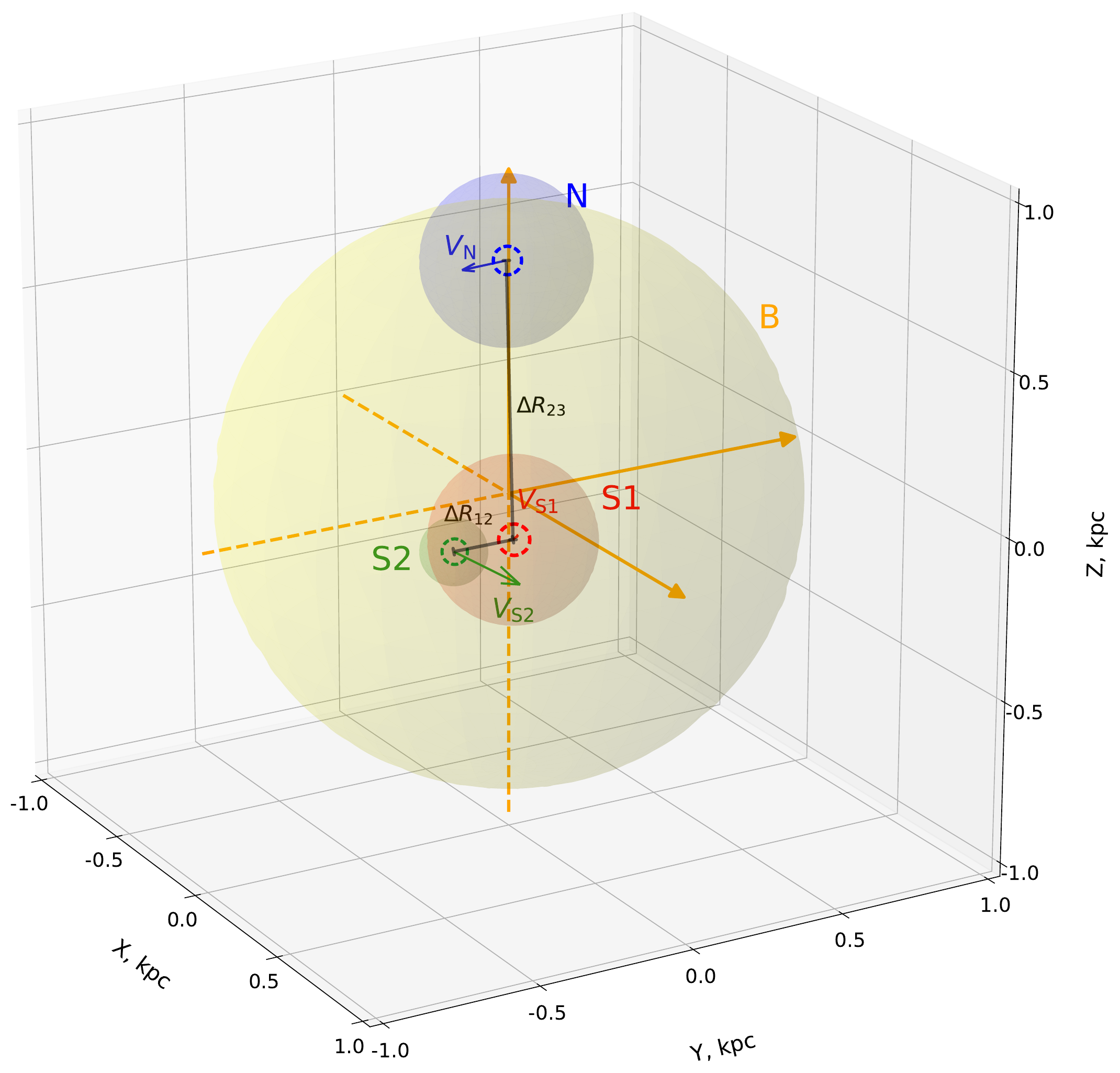}
\caption{NGC~6240 initial configuration for our triple SMBH physical model. The colour spheres represent the stellar bulges {\tt S1}, {\tt S2}, and {\tt N} with central SMBHs (circles with dash edges) and stellar bulge {\tt B} without a central black hole. SMBH initial velocities are noted $V_{\rm \tt S1,S2,N}$. Initial separation for pair {\tt S1+S2} is noted as $\Delta R_{12}$ and for pair {\tt S+N} is noted as $\Delta R_{23}$. The initial velocities are calculated in the assumption of initial orbital eccentricity $e_{0}=0.5$ for both binary pairs. The radii of the spheres are scaled according to their bulge sizes {(cutting the radius of the Plummer spheres)}.}
\label{fig:init-pos}
\end{figure}

In Fig.~\ref{fig:init-pos} we present the main physical components of our model. The masses and Plummer radii of physical components are presented in Table~\ref{tab:init-par}. The observed separation between {\tt S1} and {\tt S2} is $\Delta R_{\rm 12} = 198$~pc. The observed separation between {\tt N} and the dynamical centre of binary {\tt S1} + {\tt S2} is $\Delta R_{\rm 23} = 856$~pc, according to \cite{Kollatschny2020}. {Because these distances are the projected separations between the components, it is obvious, that these values are the lower limits of the real 3D separations between SMBHs.} To represent the large amount of gas which is placed between the {\tt S} and {\tt N} components, we added the so-called `stellar bulge' ({\tt B} -- between, in Table~\ref{tab:init-par} and yellow colour in Fig.~\ref{fig:init-pos}). 

Each component is surrounded by its bound stellar systems with simple Plummer density distribution \citep{Plummer1911} as follows:
\begin{equation}
\rho(r) = \frac{3 M_0}{4 \pi a^3} \left( 1+\frac{r^2}{a^2} \right)^{-\frac{5}{2}},
\end{equation}
which produces the cumulative mass distribution:
\begin{equation}
M(< r) = M_0 \frac{r^3}{(r^2 + a^2)^{3/2}},
\end{equation}
where $M_0$ is the total mass of each galactic bulge (total stellar mass $M_{\ast}$ in Table~\ref{tab:init-par}) and $a$ is a scale factor that characterises the size of each nucleus (Plummer radius in Table~\ref{tab:init-par}). {The $a$ scale factors for each component were calculated according with the $M_0$ masses, taking in account the central stellar densities reported in \cite{Kollatschny2020}.} Due to the flat central distribution of the Plummer profile, our estimated triple SMBH hardening will be smaller compared to the more centrally peaked core distribution profiles \citep{Jaffe1983,Hernquist1990,Dehnen1993}. Thus, using the Plummer distribution we estimated the minimum numerical hardening for our {\tt S1}~versus~{\tt S2} and {\tt S}~versus~{\tt N} components.

\subsection{Numerical realisations}\label{sec:num-real}
Based on our physical model, we constructed our numerical realisations with different total particle numbers: $N$ = 135k, 270k and 540k. For each of these numerical realisations, we created the particle distributions with initial position and velocities for five different random realisations. Thus,  we obtained a total of 15 independent numerical realisations of the same physical model (see Table~\ref{tab:init-par} and Fig.~\ref{fig:init-pos}). The dynamical evolution of one of the 540k systems is shown in the video published on YouTube\footnote{Triple SMBH evolution with the ZLK oscillation: ~\url{https://www.youtube.com/watch?v=Jy919R6f61g}}.


To scale our dynamical modelling $N$-body or H\'{e}non units\footnote{H\'{e}non units: \\~\url{https://en.wikipedia.org/wiki/N-body_units}} (NB), we used the observed physical parameters of the system; namely the distance between the southern components {\tt S1} and {\tt S2} and the total mass of the {\tt S} system. In this way, the NB units for mass, distance, velocity, and time can be respectively presented as


$M_{\rm NB} = 1.3\times 10^{{10}}\rm\;M_{\odot}$,\;\;\;\;\;\;\;\;\;\; 
$R_{\rm NB} = 198$~pc,

$V_{\rm NB} \;= 531.4$~km~s$^{-1}$,\;\;\;\;\;\;\;\;\;\;\;\;
$T_{\rm NB} = 0.364$~Myr.


For the global dynamical integration, we used the high-order parallel $N$-body code \PGPU\footnote{$N$-body code \PGPU: \\~\url{ https://github.com/berczik/phi-GPU-mole}} which is based on the fourth-order Hermite integration scheme with hierarchical individual block time steps \citep{Berczik2011, BSW2013}. The current version of the \PGPU code uses native GPU support and direct code access to the GPU using the NVIDIA native CUDA library. The current version of the code allows to have for each particle the individual softening. For the SMBH particles, we used exactly zero softening. For the SMBH versus field particle (star) interactions, we used some small effective softening: $10^{-6}\times R_{\rm NB}\approx0.2$~mpc. For the field particles self-interaction we also used a reasonably small softening parameter $10^{-4}\times R_{\rm NB}\approx20$~mpc. 

{The correct dynamical integration of the particles} can be sensitive to the global integration time step parameter, $\eta$ \citep{MA1992}. In the particular case of the fourth-order Hermite integration scheme, a time step can be written as\begin{eqnarray}
\label{eq:aarseth-timestep}
\Delta t_{i} &=& \eta \cdot \frac{A_i^{(1)}}{A_i^{(2)}}, \\
A_i^{(k)} &=& \sqrt{\absak{k-1}\absak{k+1} + \absak{k}^2},
\end{eqnarray}
where $\boldsymbol{a}^{(k)}$ is the $k^{\rm th}$ derivative of the $i$-th particle acceleration. Thus, the time steps are directly proportional to the $\eta$ parameter, which is ultimately responsible for the total integration accuracy itself. For the highest-order integration Hermite scheme, the generalised Aarseth criterion can be found in \cite{Nitadori2008}. Based on our earlier experience \citep{Sobolenko2022, Sobolenko2021}, for our set of models, we used $\eta = 0.1$ as
a general parameter. 

We also modified the standard Hermite hierarchical time step scheme, adding the extra $\eta_{\rm SMBH} = 0.05$ for the SMBH particle criteria. One more feature in our integration scheme was a minimisation and forcing synchronisation of the SMBH particle time steps. This way, we always had the SMBH particle times synchronised between each other and 
with a minimum time step between all particles.\ We already thoroughly tested and used our modified \PGPU code for a wide range of models that involve the evolution of multiple SMBHs \citep{Sedda2019, Mirza2017, Khan2018, Koehn2023}. 

Since in this case the post-Newtonian forces were not used in the simulations, we cannot talk about the physical merging of the black holes,  only about the formation/not formation of the stable bound or unbound hierarchical triple systems (HTSs). 

\section{Dynamical evolution of the triple SMBH system with the ZLK oscillation}\label{sec:triple}

The results of the formation/non-formation of SMBH HTSs are presented for three numerical models with particle numbers 135k, 270k, and 540k. Each of them had five different randomisation seeds for the initial positions and velocities of particles. As an illustration of the dynamical evolution of the system, we present a video of the density distribution for the highest particle number, 540k, on YouTube\footnote{Triple SMBH evolution with the ZLK oscillation: ~\url{https://www.youtube.com/watch?v=Jy919R6f61g}}.

A summary of the results of modelling the SMBH dynamics in the NGC~6240 galaxy are presented in Table~\ref{tab:result}. As we see, we have two separate physical phases of system evolution. During the first phase of evolution, the SMBHs {\tt S1} and {\tt S2} become bound. This event, independently of particle numbers, happens around $T_{\rm b, {\tt S1+S2}}\approx3.6$~Myr from the beginning of the integration. The second phase started when the {\tt N} component became bound with the {\tt S1+S2} system, hereafter {\tt S}. This happens around $T_{\rm b, {\tt S+N}}\approx18$~Myr from the beginning of integration, again independently of the particle numbers.  

\begin{table}
\caption{Simulation results of the formation or non-formation of hierarchical triple SMBH systems and follow-up ZLK oscillations.}
\label{tab:result}
\centering
\renewcommand{\arraystretch}{1.2}
\begin{tabular}{ccccc}
\hline
\hline
$N$ & $T_{\rm b,{\tt S1+S2}}$ &  $T_{\rm b, {\tt S+N}}$ & HTS & ZLK \\
(1) & (2) & (3) & (4) & (5) \\
\hline
\hline
135k &  &  & 4 from 5 & 4 from 5 \\
270k & 3.6 & 18 & 4 from 5 & 4 from 5 \\
540k &  &  & 5 from 5 & 5 from 5 \\
\hline
\end{tabular}
\begin{minipage}{\linewidth}
\smallskip
\textit{Note:}
(1) -- particles number, (2) -- formation time of the bound system {\tt S1+S2} in Myr, (3) -- formation time of the bound system {\tt S+N} in Myr, where the {\tt S} is a centre of {\tt S1+S2} mass system, (4) -- number of systems that form a HTS, (5) -- number of systems with the ZLK oscillation.
\end{minipage}
\end{table}

In Fig.~\ref{fig:r-evol}, as an illustration, we present the evolution of separation between SMBH pairs for five realisations of the numerical model with 540k particles: for {\tt S1+S2} and {\tt S+N}. As a guiding line, we plot a level of ten times the sum of the three SMBH Schwarzschild radii. We can see that after the third SMBH ({\tt N} component) arrives at the {\tt S} component and becomes bound with {\tt S}, there is a significant oscillation of $\Delta  R_{\tt S1+S2}$. One of the simplest explanations of this behaviour is a possible manifestation of the ZLK mechanism \citep[][for for a historical overview and nomenclature, we refer to \citealt{Ito2019}]{Zeipel1910, Lidov1961, Lidov1962, Kozai1962}.

{The system dynamical evolution for the other two field particles, numbers 135k and 270k, is also presented in Appendix \ref{app:num_ptcl}.} We see the HTS formation and the subsequent ZLK oscillations only in four cases out of the five numerical randomisations in the case of the 135k and 270k particle numbers. For the highest particle number, 540k, for all initial randomisation seeds, the HTS and ZLK mechanisms are observed. From this simple fact, we can conclude that in the case of a large enough particle number, the HTS and ZLK formation is inevitable in our initial model configuration. 

{In Fig.~\ref{fig:r-evol}, at the late phase of the dynamical evolution for the randomisations 1 (blue) and 2 (green) we see a classical ejection effect in these HTSs. The lowest-mass SMBH ({\tt S2}) during the end phase of the simulation was ejected from the central region. }

In Fig.~\ref{fig:e-evol}, we present the evolution of eccentricity $e$ of the SMBH pairs for five realisations of the numerical model with 540k particles: for {\tt S1+S2} and {\tt S+N}. As we would expect due to the ZLK mechanism, we can see the significant oscillation of the {\tt S1+S2} binary SMBH eccentricity after the formation of bound {\tt S} and {\tt N} systems. The exact range of eccentricity oscillations depends on the particle randomisation, but in all cases, it is very significant and   reaches nearly $\approx1$ for each case.

The evolution of the pericenter, $r_{\rm p}$ (Fig.~\ref{fig:rp-evol}), also shows the periodicity for inner {\tt S1+S2} binary. The inner binary pericenter oscillations are mainly defined by the eccentricity oscillations. {The inverse semimajor axis of the {\tt S1+S2} binary during the ZLK phase of simulation is extending almost linearly  (i.e. the semimajor axis of the inner binary is constantly shrinking)}; however, this happens on a much longer timescale compared to the $r_{\rm p}$ oscillation period, which clearly reflects the eccentricity oscillations of the inner binary (see Fig.~\ref{fig:e-evol}).

In Fig.~\ref{fig:i-evol}, we present the mutual inclination angle between the angular momentum of the inner {\tt S1+S2} and outer {\tt S+N} binaries \citep[see Fig.~1]{Naoz2016}. As we can see, the inclination angle evolution is in very good agreement with the expected theoretical limits of the Kozai angles: $\cos i_{\rm min}$ = $\pm \sqrt{3/5} \approx39.2^\circ$.

Here, we can note that our triple SMBH behaviour is more complex compared to the pure three-body ZLK effect. Together with the three SMBHs, our dynamical system contains a few hundred thousand field particles (see the video on YouTube\footnote{Triple SMBH evolution with the ZLK oscillation: ~\url{https://www.youtube.com/watch?v=Jy919R6f61g}}). Of course, the presented above spatial configuration is only one of the possible representations of the observed data set. In the present work, for simplicity and a quick set of the modelling, we limited ourselves to this single interpretation. In some sense, the chosen physical model works in favour of the ZLK oscillations in the triple SMBH system and it is enough to demonstrate this possible effect.

{As an additional run for 540k in the first randomisation, we generated the system with a smaller initial orbital inclination angle ($\approx$75$^\circ$ instead of 90$^\circ$) and with a slightly larger initial orbital eccentricity $\approx$0.6 instead of 0.5 of the {\tt S+N} outer binary. The results of this extra run is presented in the top-right panel (with orange colour) in Figs.~\ref{fig:r-evol}-\ref{fig:i-evol}.}

For a detailed analysis of the ZLK oscillations in our triple systems, we ran separate extremely high time-resolution simulations with the data output time interval of $\approx1.6$~year. These simulations were run for the 540k for two intervals of the time around 18~Myr (in a moment of ZLK formation) and around 22~Myr (see middle and right panels in Fig.~\ref{fig:zlk-meh}). These time intervals are also marked as vertical black dotted lines in the left panels in Fig.~\ref{fig:zlk-meh}. Here, we see the stable ZLK effect with the almost constant periodicity and inclination angle amplitude.  


{As an extra test of the possible {\tt S1+S2} binary post-Newtonian merger attributed to the high orbital eccentricity of the ZLK effect, we started a set of restricted three-body SMBH simulations, including the combination of different post-Newtonian terms (PN0 - Newtonian, PN1~$\sim1/c^2$, PN2~$\sim1/c^4$, and PN2.5~$\sim1/c^5$). The resulting plots of the eccentricity evolution of the different models are presented in Fig.~\ref{fig:ecc-pn}. The initial positions and velocities were chosen from the full scale $N$-body model, with a $N$=540k first randomisation at three moments of time in the first ZLK oscillation period (around $\sim$18.5 Myr), namely: when orbital eccentricity for the inner binary {\tt S1+S2} was 0.5, 0.7, and 0.9. 

We individually turned on PN0 (`Newtonian'), PN0+PN1, PN0+PN1+PN2, only PN0+PN2.5, and PN-full = PN0+PN1+PN2+PN2.5 (i.e. with `full') terms for three-body post-Newtonian code \citep{Sobolenko2017, Sobolenko2022}. As we can see, the post-Newtonian terms (mainly the PN1) significantly change the whole ZLK effect evolution. With the PN-Full regime, we see the possible gravitational wave merger only for a very high initial eccentricity ($\gtrsim$0.9). The starting points with the lower initial eccentricities (0.5 and 0.7) simply show the non-growing, fixed eccentricity, evolution (i.e. basically no ZLK effect), with no subsequent gravitational wave merging for the {\tt S1+S2} inner binary in NGC~6240. In this sense, our set of restricted three-body post-Newtonian simulations support the ideas already presented in the works of \cite{Mannerkoski2021}, \cite{Bonetti2016}, and \cite{Blaes2002} -- ultimately demonstrating that the PN1 term may suppress the ZLK effects. }

\section{Conclusions}\label{sec:conc}

We carried out a dynamical modelling of the evolution for the triple SMBH system in the dense stellar environment in the core of the NGC~6240 galaxy. Our physical model was constrained based on the latest available observational data \citep{Kollatschny2020} and information from the literature \citep{Komossa2003, Fabbiano2020} for this object. We run a set of numerical simulations using the well-tested dynamical $N$-body \PGPU code. Based on the general physical model, we constructed three numerical models with different numbers of particles (135k, 270k, and 540k), where each model had five independent sets of randomisations. 

According to our simulation results, for all models and randomisations, the bound system from {\tt S1+S2} components formed at $\approx3.6$~Myr. The formation of a HTS triple bound system {\tt S+N} occurs at $\approx18$~Myr. During the {subsequent pure-Newtonian} simulation of the evolution of triple SMBHs and the surrounding environment in NGC~6240, a ZLK oscillation was detected for the binary components {\tt S} (the centre of mass of the {\tt S1+S2}) and {\tt N} systems. {For these runs,} the inclination angle between the orbital angular momentum of binary components aptly coincides with the theoretical calculations of the ZLK mechanism. 

{In contrast, our extra few-body post-Newtonian runs (for one randomisation case) show it is only for the large inner binary initial eccentricity (in our case $\gtrsim$0.9) that we have a possibility of the inner binary merging due to the post-Newtonian energy radiation effects. For the lower eccentricity cases, the test runs show no sign of any possible merging and no hint of any ZLK oscillations in the system.}

\begin{acknowledgements}

{The authors thank the anonymous referee for a very constructive report and suggestions that helped significantly improve the quality of the manuscript.}

The PB and MS thanks the support from the special program of the Polish Academy of Sciences and the U.S. National Academy of Sciences under the Long-term program to support Ukrainian research teams grant No.~PAN.BFB.S.BWZ.329.022.2023.

PB and MI acknowledge the support of the  Science Committee of the Ministry of Education and Science of the Republic of Kazakhstan (Grant No.~AP14870501).

MS acknowledges the support under the Fellowship of the President of Ukraine for young scientists 2022-2024.

PB and MS gratefully acknowledge the Gauss Centre for Supercomputing e.V. (www.gauss-centre.eu) for funding this project by providing computing time through the John von Neumann Institute for Computing (NIC) on the GCS Supercomputer JUWELS at Julich Supercomputing Centre (JSC).

\end{acknowledgements}

\bibliographystyle{mnras}  
\bibliography{NGC6240-triple}   

\onecolumn
\begin{appendix}
\section{Evolution of the orbital parameters for the runs}\label{app:num_ptcl}

\begin{figure*}
\centering
\includegraphics[width=0.79\linewidth]{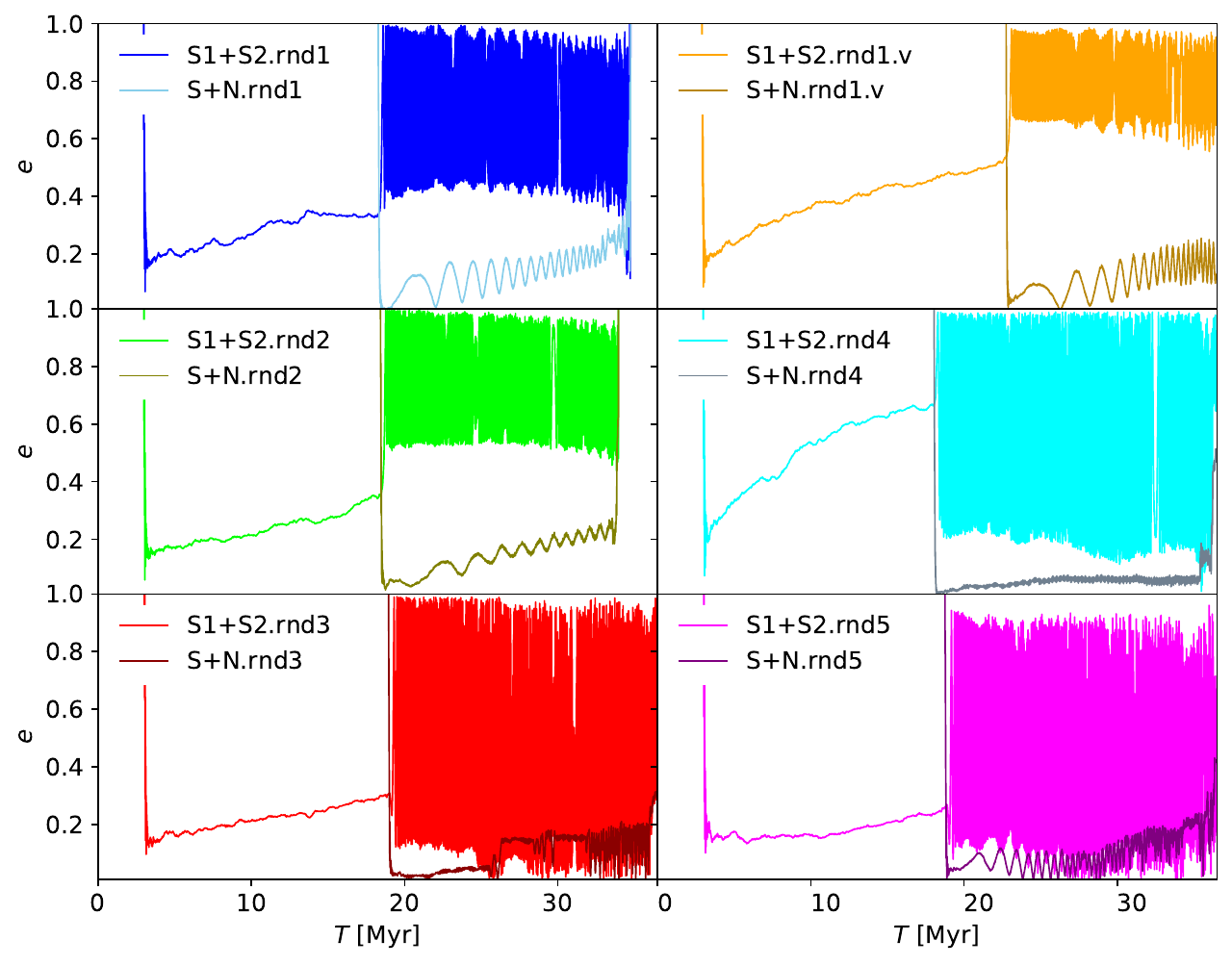}
\caption{Evolution of eccentricity $e$ for {\tt S1+S2} and {\tt S+N} systems for numerical model with particle number 540k and for the same five different randomisation seeds. }
\label{fig:e-evol}
\end{figure*}

\begin{figure*}
\centering
\includegraphics[width=0.79\linewidth]{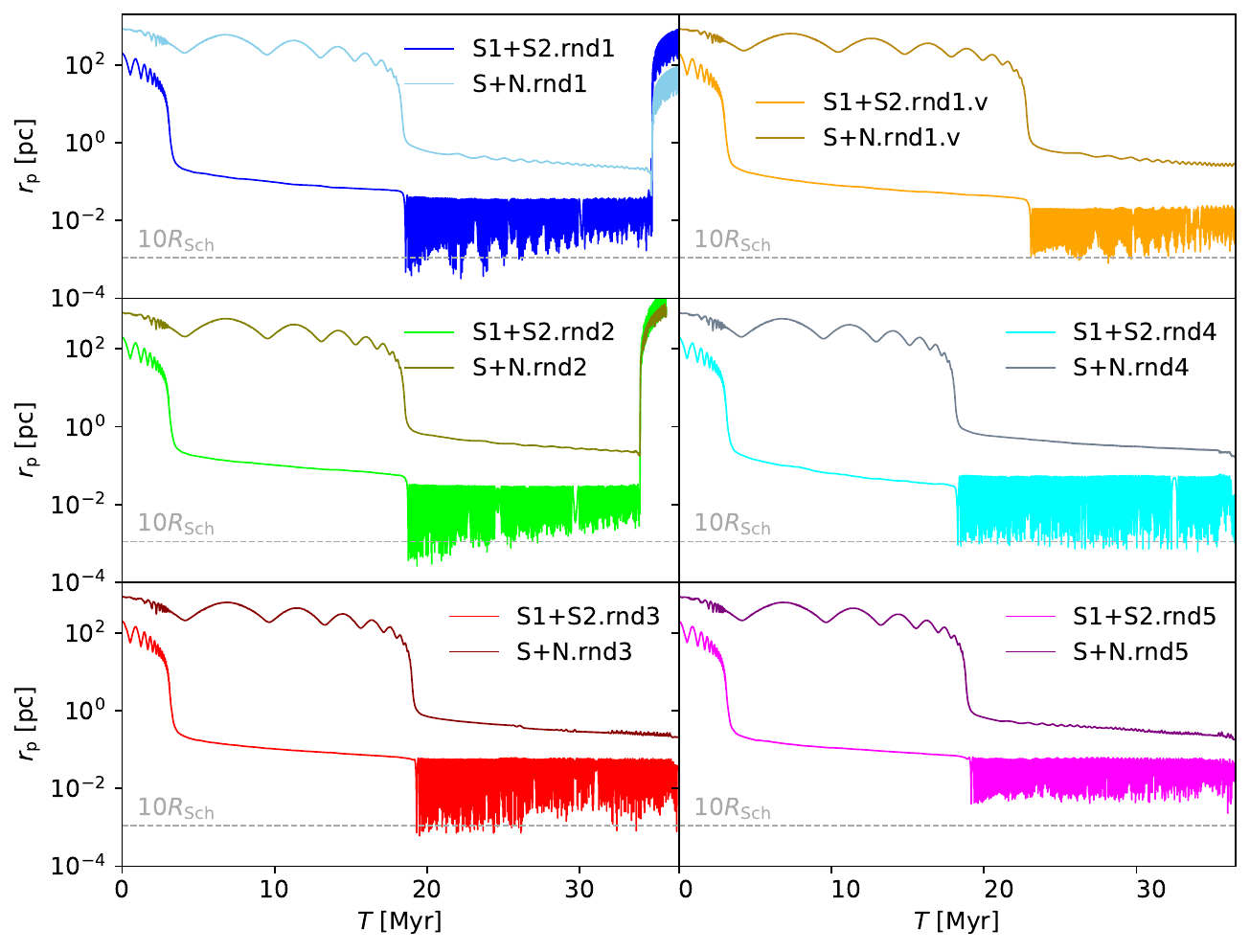}
\caption{Evolution of pericenter, $r_{\rm p}$m for {\tt S1+S2} and {\tt S+N} systems for a numerical model with the particle number 540k and for the five different randomisation seeds. The dotted grey line is the level of the ten Schwarzschild radii (as in Fig.~\ref{fig:r-evol})}
\label{fig:rp-evol}
\end{figure*}

\begin{figure*}
\centering
\includegraphics[width=0.79\linewidth]{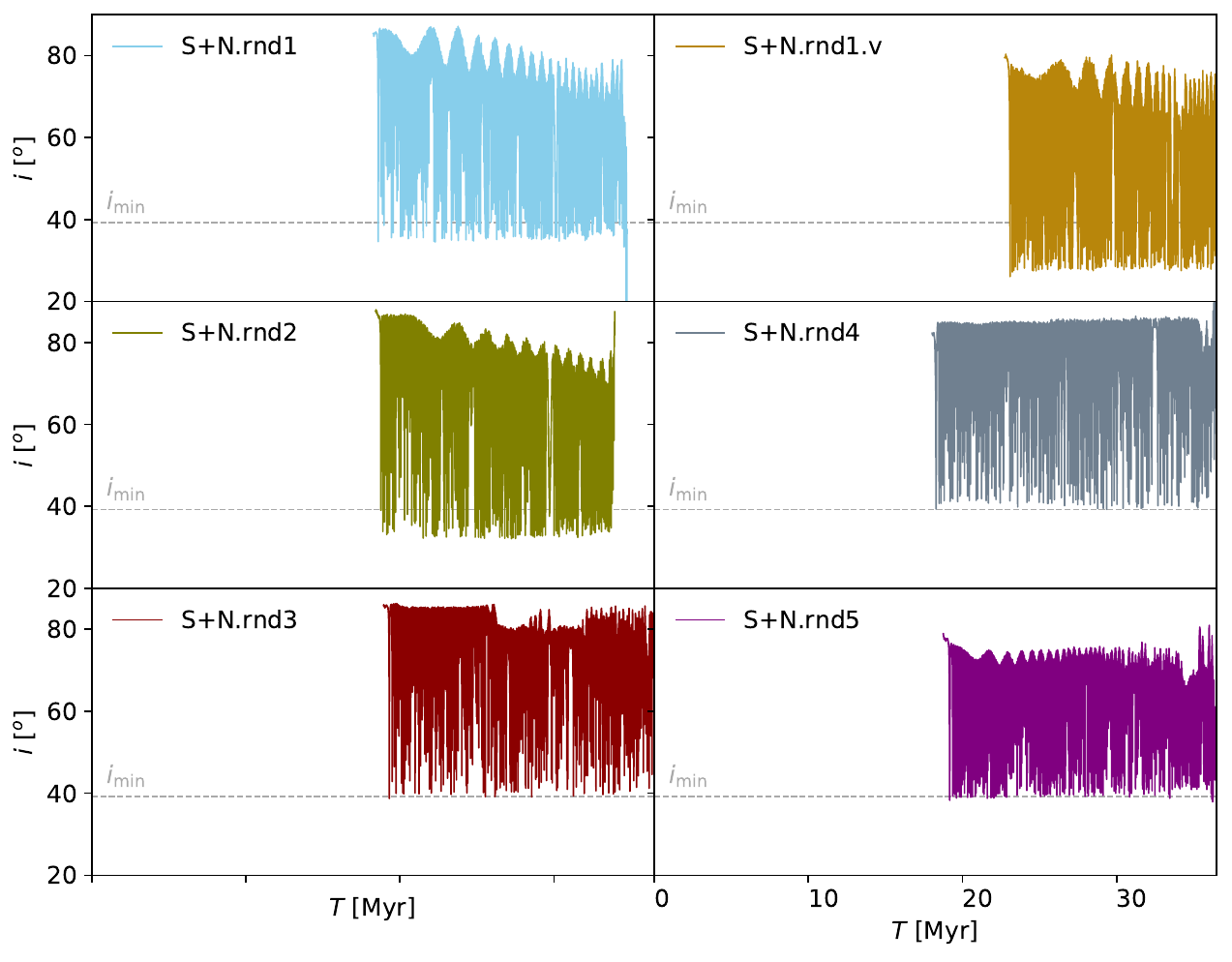}
\caption{Evolution (ZLK oscillation) of the angle, $i,$ between the {\tt S1+S2} and {\tt S+N} SMBH binary systems for a numerical model with particle number 540k and for the five different randomisation seeds. The dotted grey lines represent the limiting theoretical Kozai angles.}
\label{fig:i-evol}
\end{figure*}

\begin{figure*}
\centering
\includegraphics[width=0.78\linewidth]{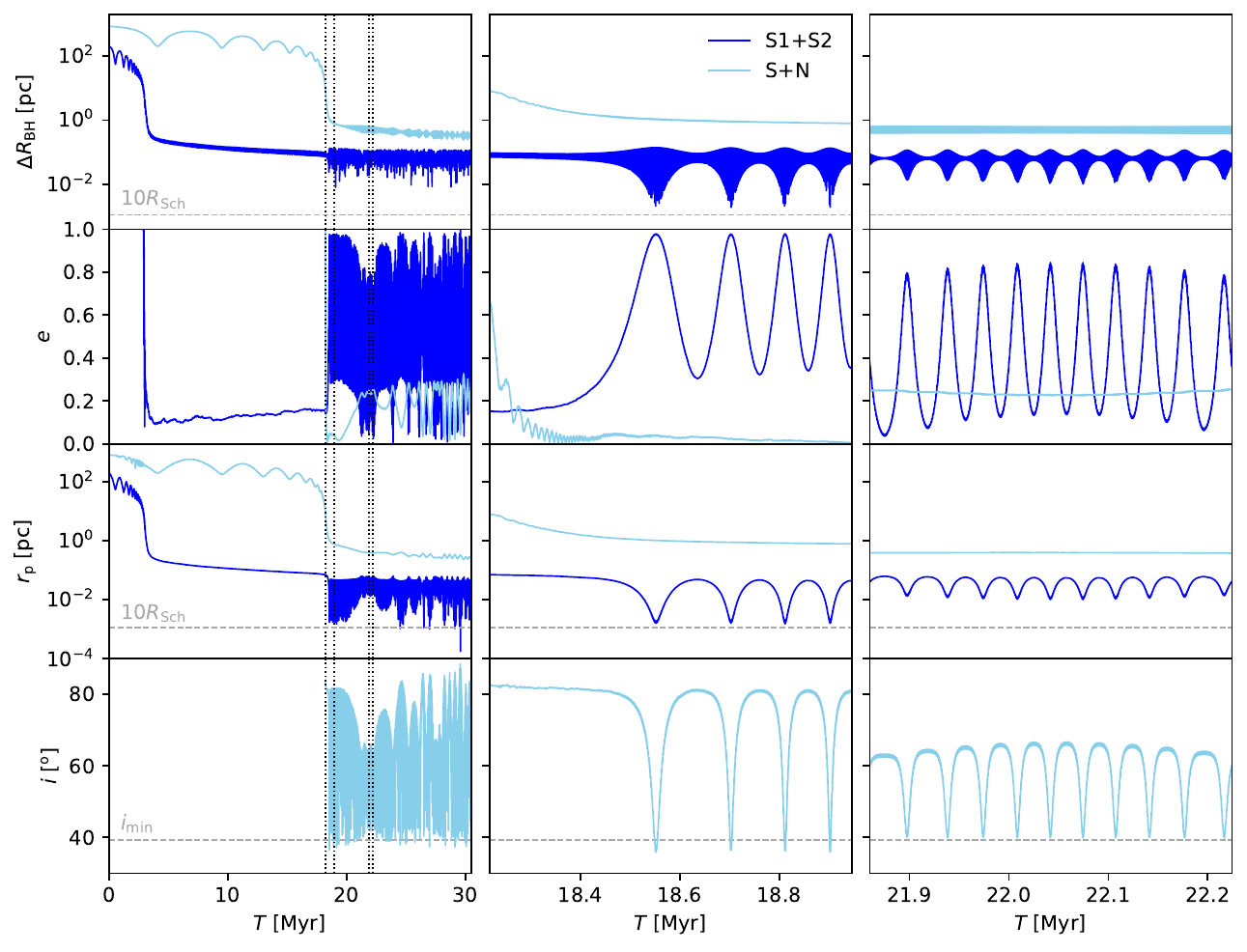}
\caption{Overall detail dynamical evolution of the selected 540k model. 
\textit{Left column:}  Long evolution till $\approx30$~Myr, where the vertical dotted black lines show the two intervals of the time for the further frequent output runs. \textit{Middle and right columns:} Higher time resolution views of the ZLK effect for the simulations. Blue and light blue lines represents the parameters evolution for {\tt S1+S2} and {\tt S+N} SMBHs from top to bottom: separation, $\Delta R$, eccentricity, $e$, pericenter, $r_{\rm p}$, angle, $i,$ between the {\tt S1+S2} and {\tt S+N} systems. On the separation and pericenter plots, the dashed grey lines represent ten Schwarzschild radii for SMBHs with summarised mass. On the angle plot, the dashed grey line represents the theoretical limiting Kozai angles, $i_{\rm min}$.}
\label{fig:zlk-meh}
\end{figure*}

\begin{figure*}
\centering
\includegraphics[width=0.79\linewidth]{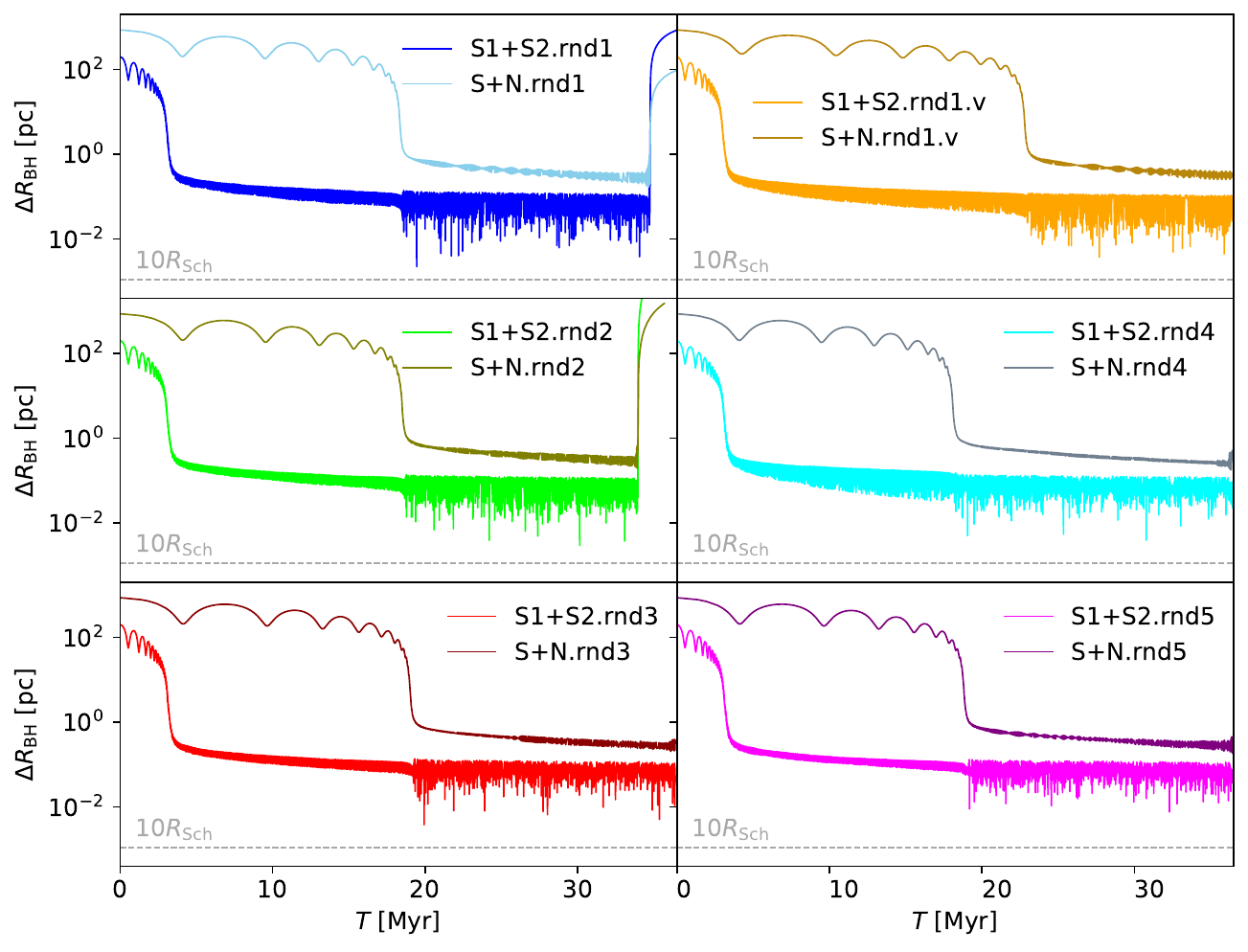}
\caption{Evolution of the separations, $\Delta R$, between SMBHs in pairs {\tt S1+S2} and {\tt S+N} for the numerical model with particle number 540k. Five plots show five different initial randomisation seeds for the particles distributions. The dotted grey line is the level of the ten Schwarzschild radii for SMBHs with summarised mass: $M=M_{\rm \tt S1}+M_{\rm \tt S2}+M_{\rm \tt N}$.}
\label{fig:r-evol}
\end{figure*}

\begin{figure}[!b]
\centering
\includegraphics[width=0.70\linewidth]{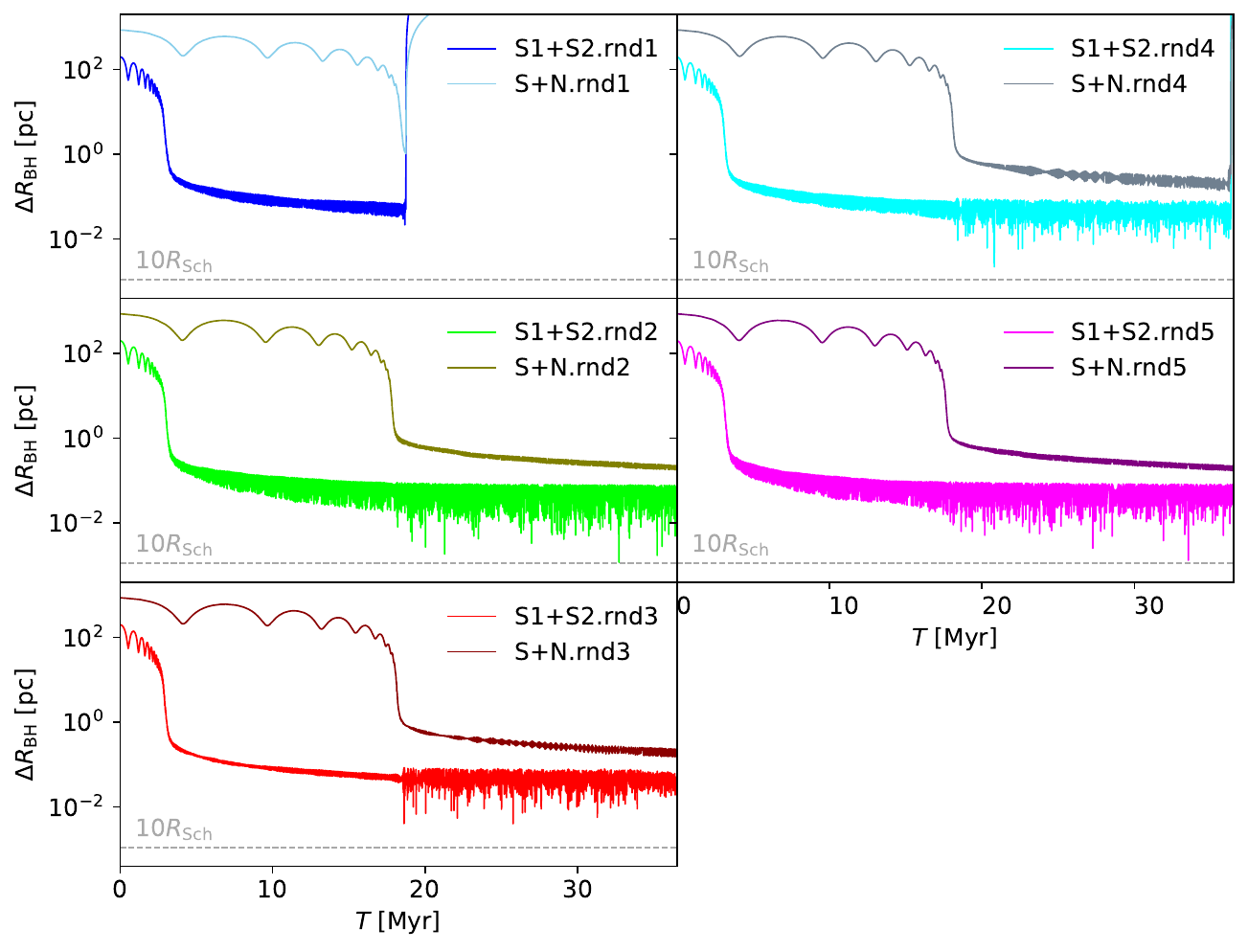}
\caption{Evolution of separations $\Delta R$ between SMBHs in pairs {\tt S1+S2} and {\tt S+N} for numerical model with particle number 135k. Five plots show five different initial randomisation seeds for the particles distributions. The dotted grey line is the level of the ten Schwarzschild radii for SMBHs with summarised masses $M=M_{\rm \tt S1}+M_{\rm \tt S2}+M_{\rm \tt N}$.}
\label{fig:r-evol-135k}
\end{figure}

\begin{figure}[!b]
\centering
\includegraphics[width=0.70\linewidth]{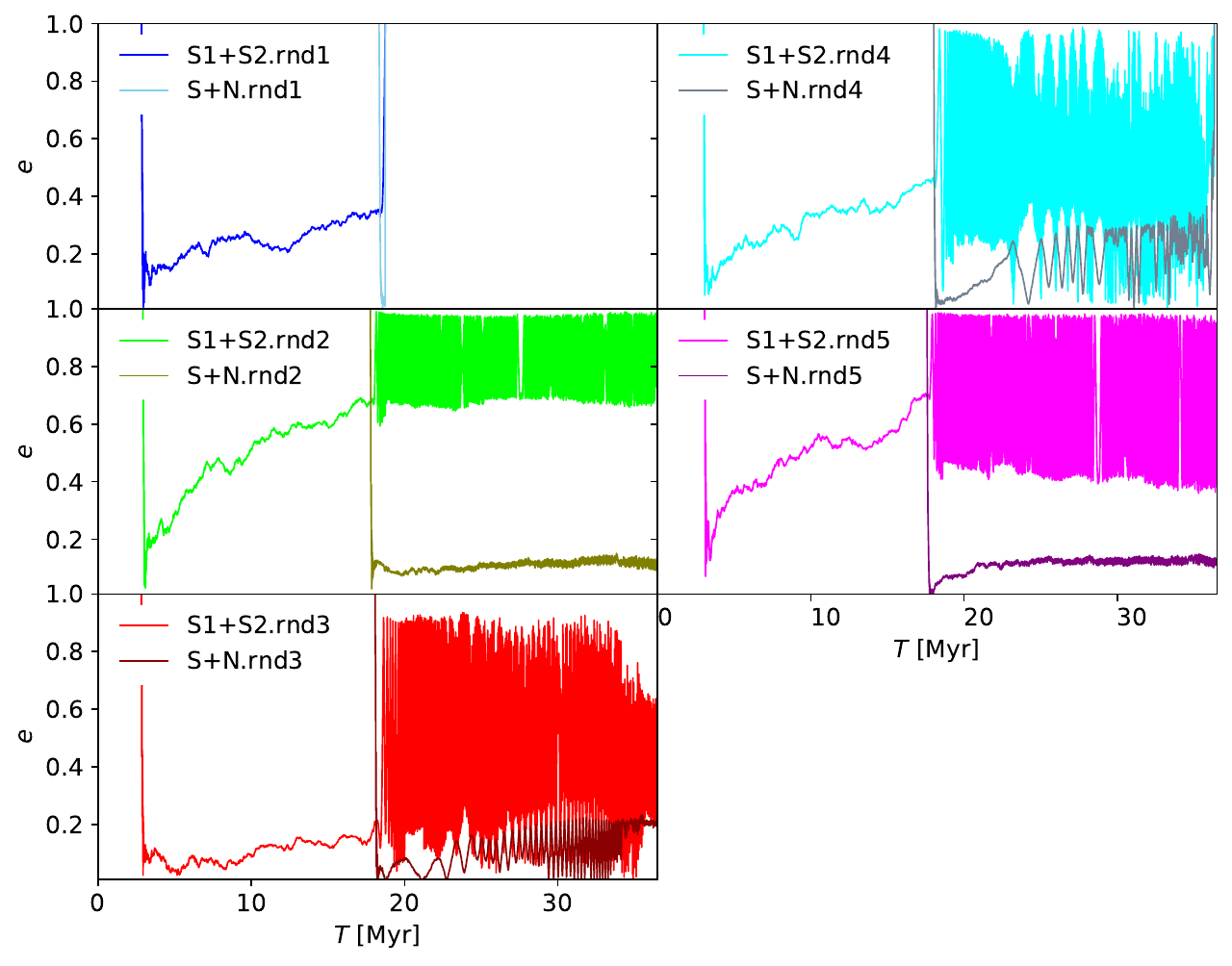}
\caption{Evolution of eccentricity, $e$, for {\tt S1+S2} and {\tt S+N} systems for numerical model with particle number 135k and for the same five different randomisation seeds.}
\label{fig:e-evol-135k}
\end{figure}

\begin{figure}
\centering
\includegraphics[width=0.79\linewidth]{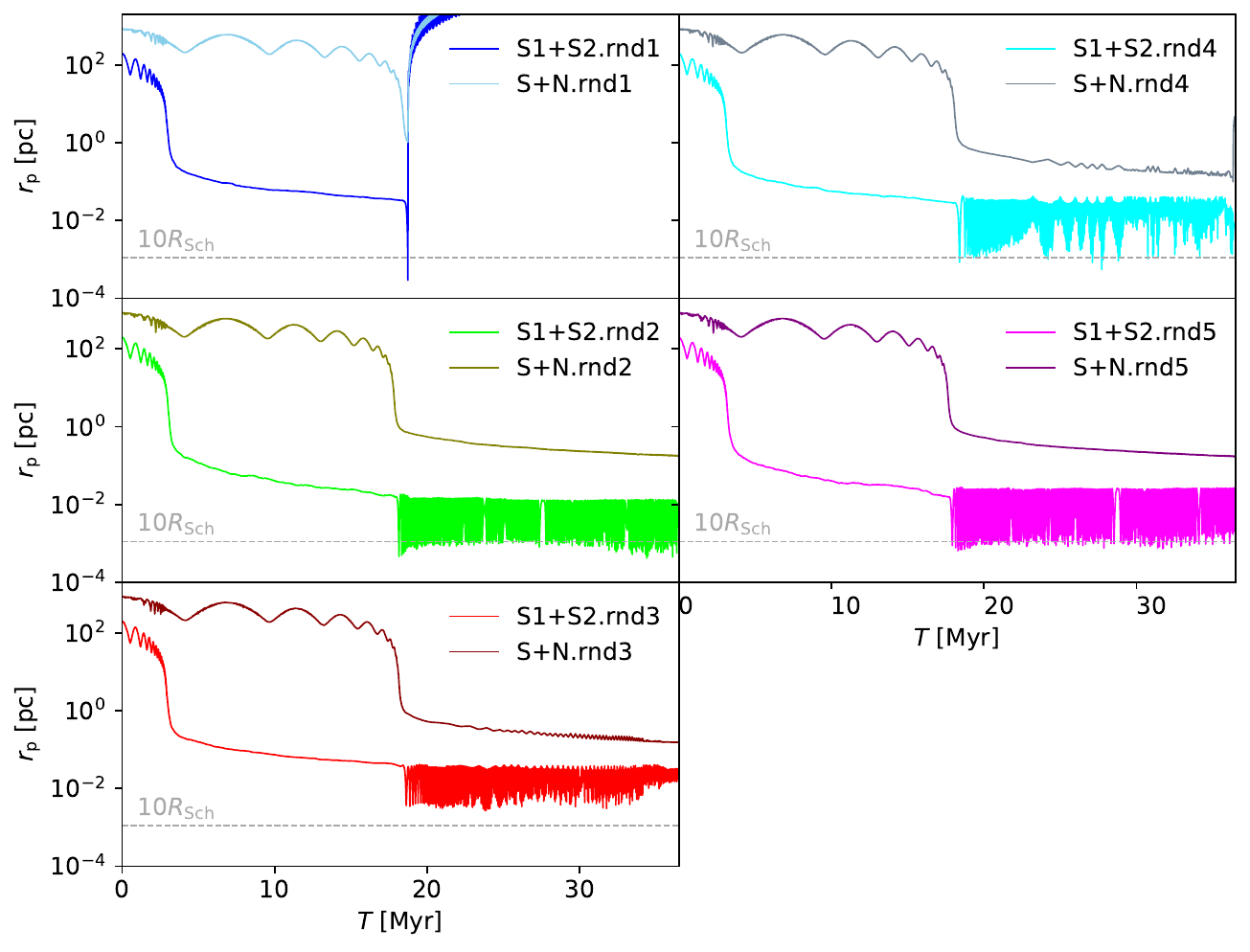}
\caption{Evolution of pericenter $r_{\rm p}$ for {\tt S1+S2} and {\tt S+N} systems for numerical model with the particle number 135k and for the five different randomisation seeds. The dotted grey line is the level of the ten Schwarzschild radii (as in Fig.~\ref{fig:r-evol})}
\label{fig:rp-evol-135k}
\end{figure}

\begin{figure}
\centering
\includegraphics[width=0.79\linewidth]{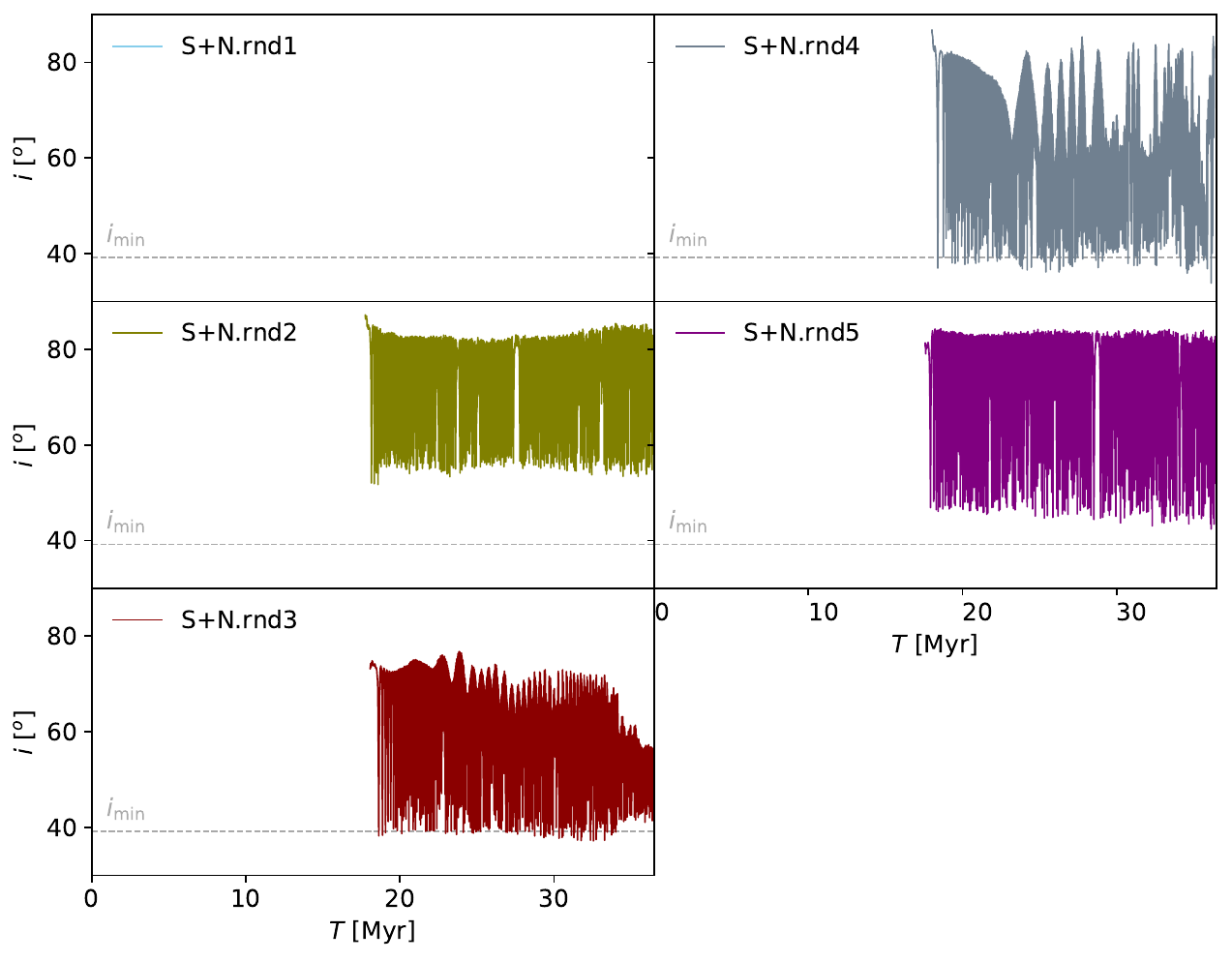}
\caption{Evolution (ZLK oscillation) of the angle $i$ between {\tt S1+S2} and {\tt S+N} SMBH binary system for numerical model with particle number 135k and for the five different randomisation seeds. The dotted grey lines represent the limiting theoretical Kozai angles.}
\label{fig:i-evol-135k}
\end{figure}

\begin{figure}[!b]
\centering
\includegraphics[width=0.70\linewidth]{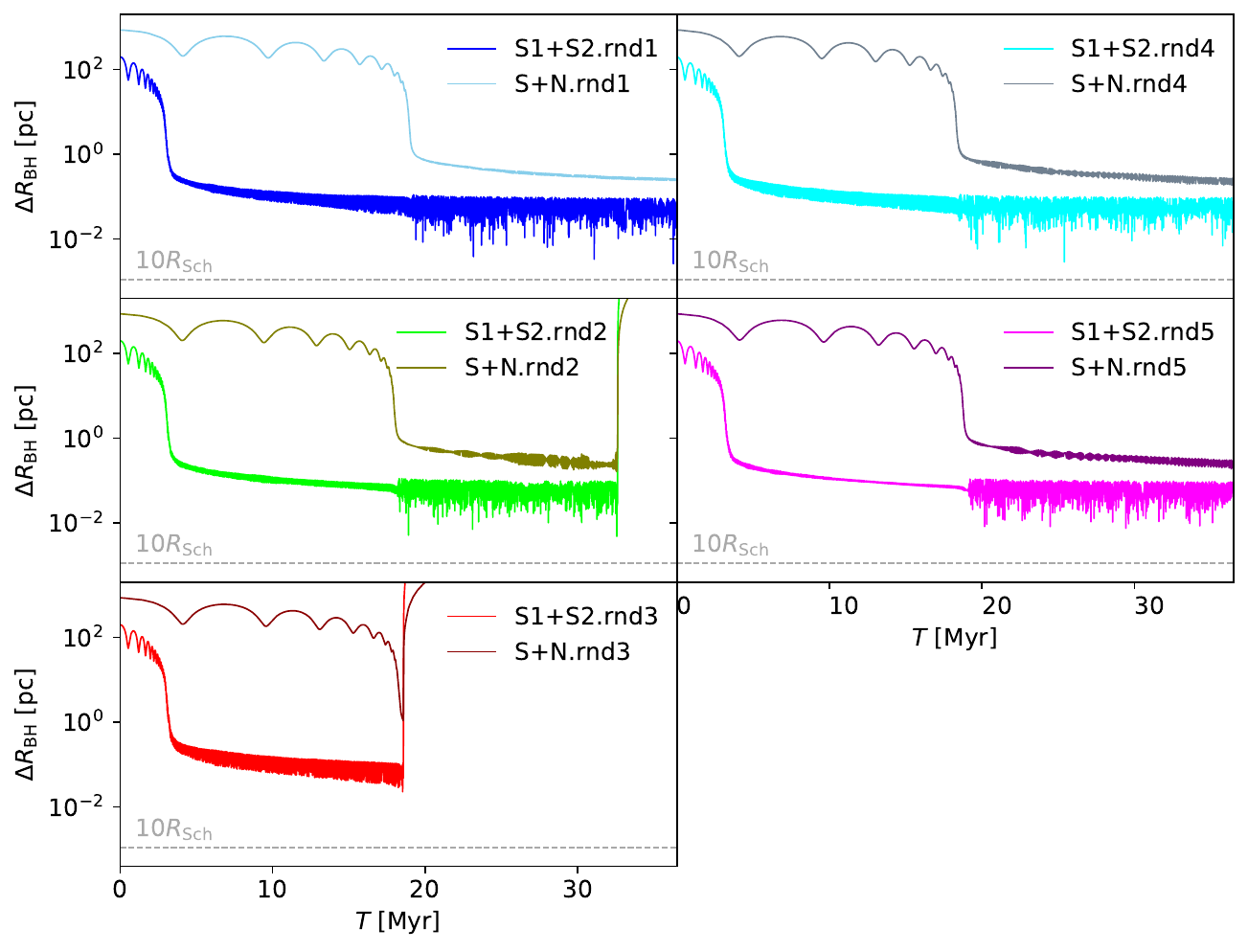}
\caption{Same as Fig.~\ref{fig:r-evol-135k} but for $N$=270k.}
\label{fig:r-evol-270k}
\end{figure}

\begin{figure}[!b]
\centering
\includegraphics[width=0.70\linewidth]{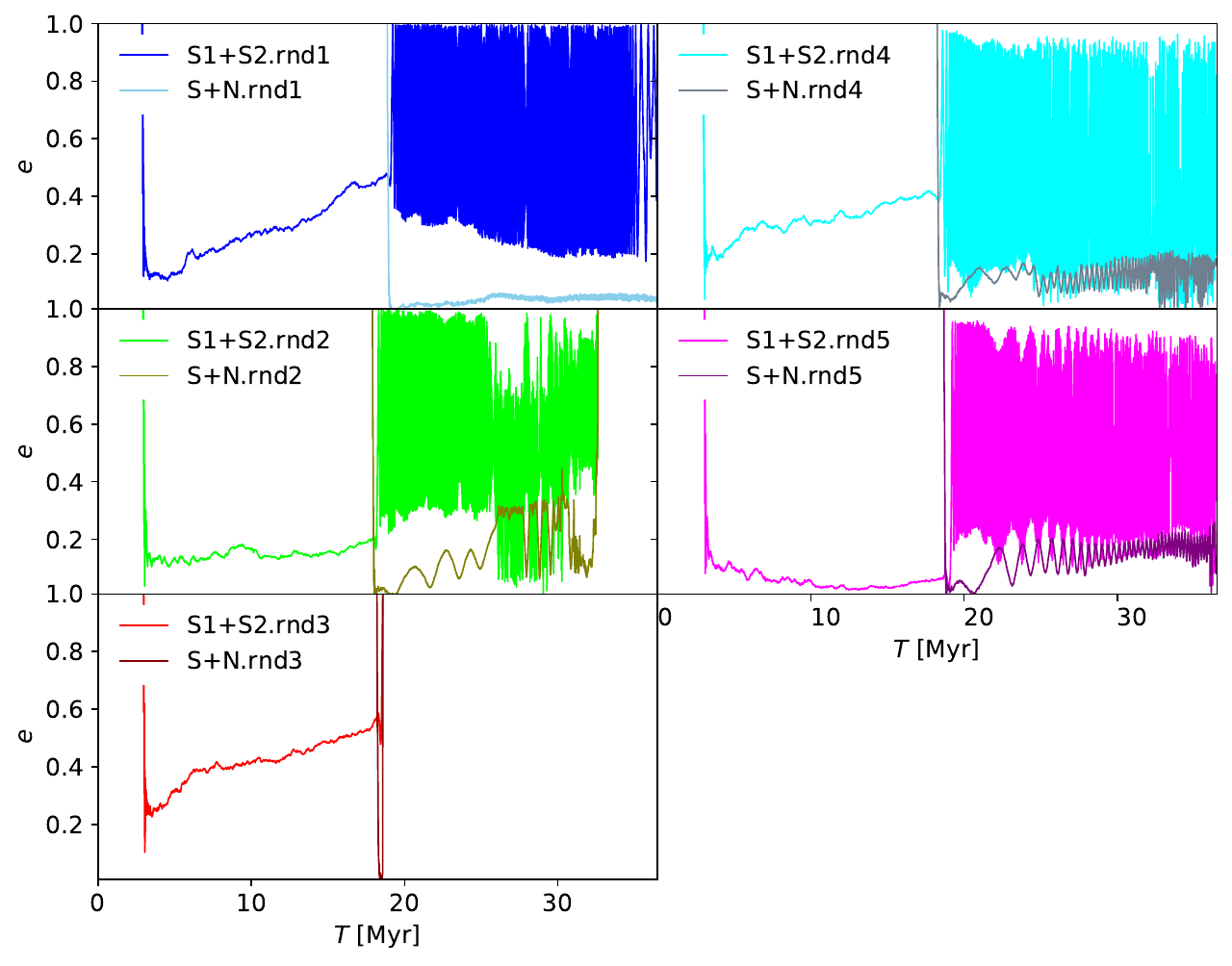}
\caption{Same as Fig.~\ref{fig:e-evol-135k} but for $N$=270k.}
\label{fig:e-evol-270k}
\end{figure}

\begin{figure}
\centering
\includegraphics[width=0.79\linewidth]{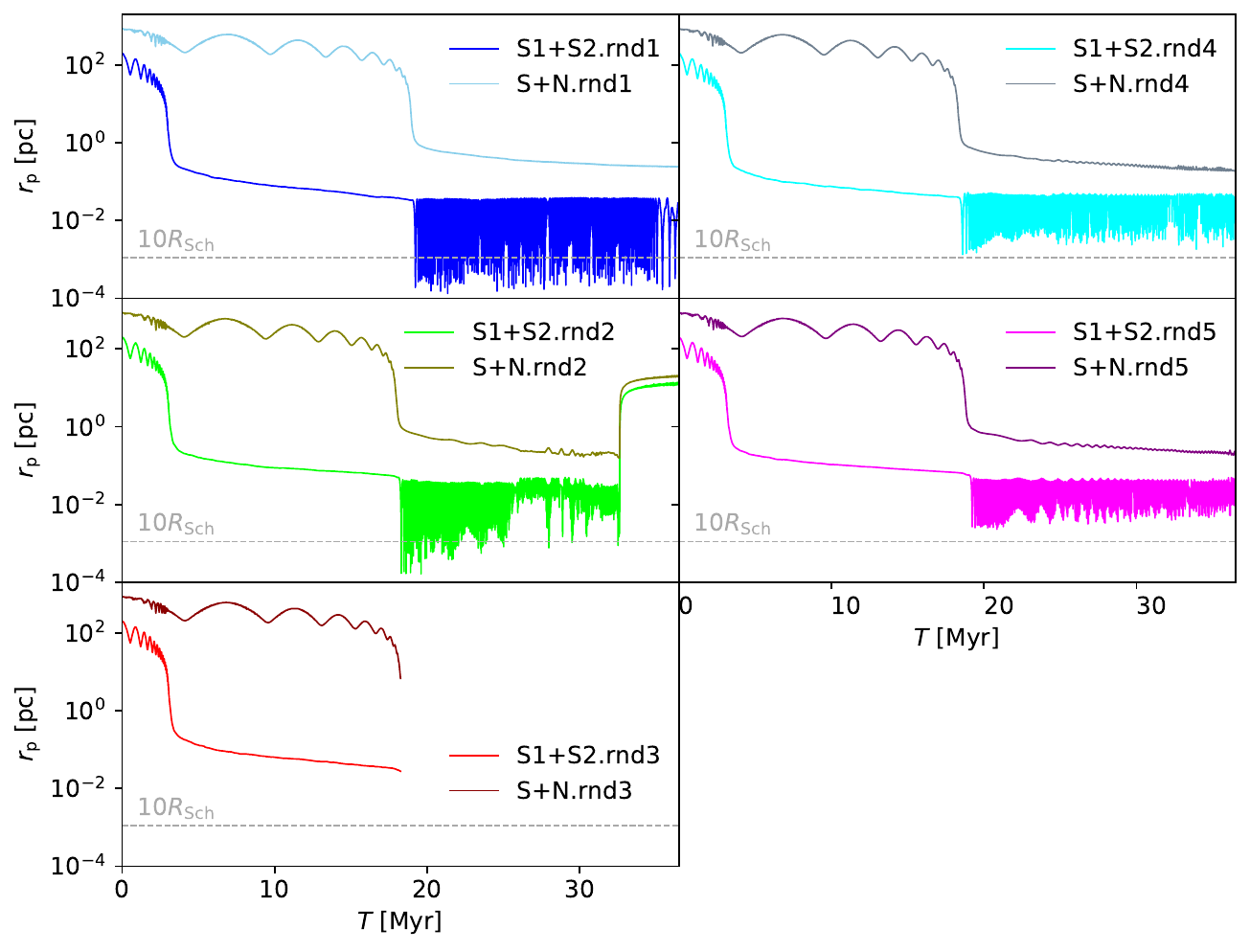}
\caption{Same as Fig.~\ref{fig:rp-evol-135k} but for $N$=270k.}
\label{fig:rp-evol-270k}
\end{figure}

\begin{figure}
\centering
\includegraphics[width=0.79\linewidth]{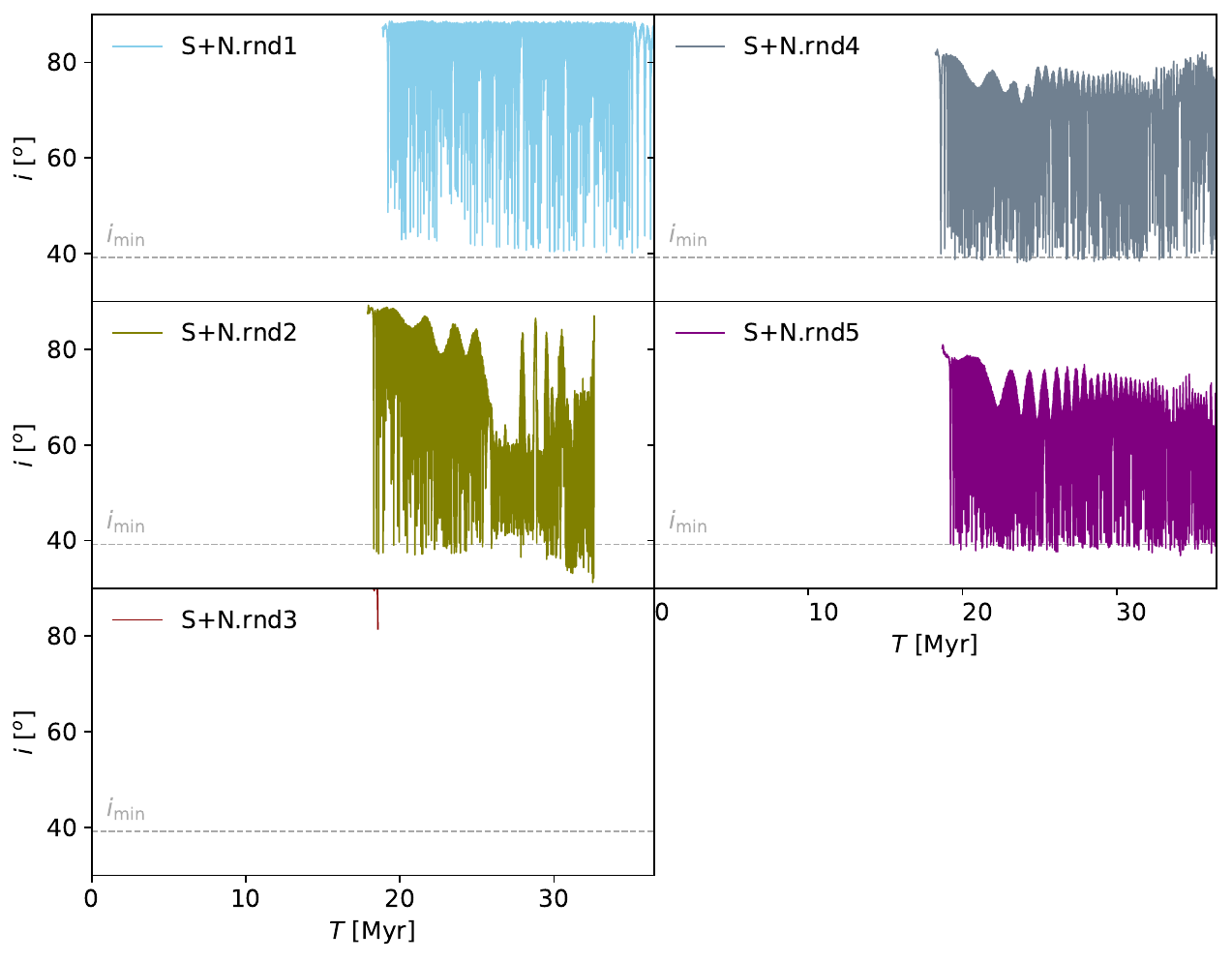}
\caption{Same as Fig.~\ref{fig:i-evol-135k} but for $N$=270k.}
\label{fig:i-evol-270k}
\end{figure}
\clearpage
\section{Evolution of the orbital parameters for the runs with PN terms}\label{app:pn-terms}

\begin{figure*}[h]
\centering
\includegraphics[width=0.52\linewidth]{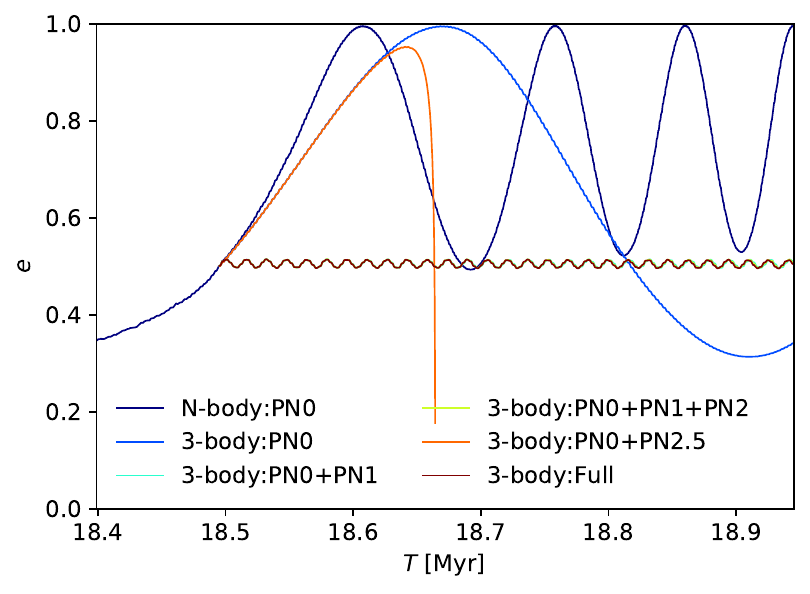}
\includegraphics[width=0.52\linewidth]{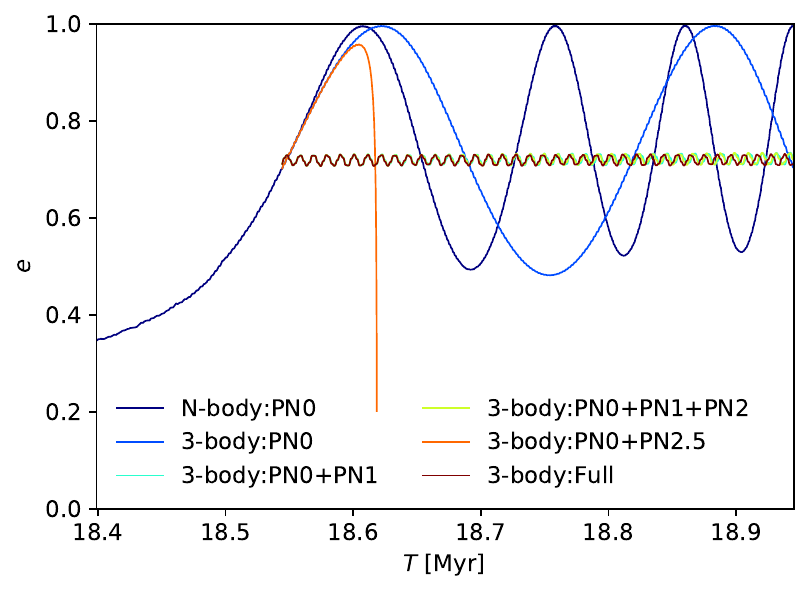}
\includegraphics[width=0.52\linewidth]{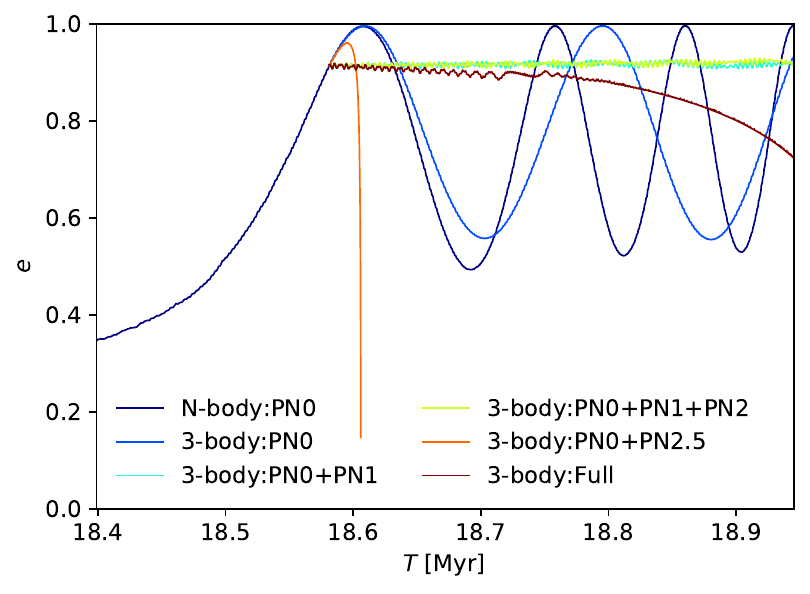}
\caption{Evolution of the eccentricity for the 3-body PN integration model. Initial positions and velocities were chosen from the full scale $N$-body model with $N$=540k first randomisation at three moments of time in the first ZLK oscillation period \textit{from top to bottom}: when orbital eccentricity for the inner binary {\tt S1+S2} was 0.5, 0.7 and 0.9. We individually turned ``on'' the different combination of PN terms in our 3-body PN code: PN0 (``Newtonian''), PN0+PN1, PN0+PN1+PN2, only PN0+PN2.5 and PN-Full = PN0+PN1+PN2+PN2.5 (i.e. with ``Full'').}
\label{fig:ecc-pn}
\end{figure*}

\end{appendix}

\end{document}